\documentclass[twocolumn,aps,preprintnumbers,amsmath,amssymb]{revtex4}

\usepackage{graphicx}
\usepackage{dcolumn}
\usepackage{bm}
\usepackage{color}

\begin{document}


\title{Anisotropic magnetic-field response of quantum critical fluctuations\\ in Ni-doped CeCoIn$_{\bm{5}}$}
\author{Makoto~Yokoyama}
\email[]{makoto.yokoyama.sci@vc.ibaraki.ac.jp}
\author{Kohei~Suzuki}
\email[]{13s2017t@gmail.com}
\affiliation{Faculty of Science, Ibaraki University, Mito, Ibaraki 310-8512, Japan\\
and Institute of Quantum Beam Science, Ibaraki University, Mito, Ibaraki 310-8512, Japan}
\author{Kenichi~Tenya}
\affiliation{Faculty of Education, Shinshu University, Nagano 380-8544, Japan}
\author{Shota~Nakamura}
\altaffiliation{present address: Department of Physical Science and Engineering, Nagoya Institute of Technology, Nagoya 466-8555, Japan}
\author{Yohei~Kono}
\author{Shunichiro~Kittaka}
\author{Toshiro~Sakakibara}
\affiliation{Institute for Solid State Physics, The University of Tokyo, Kashiwa, Chiba 277-8581, Japan}

\date{\today}
             
\begin{abstract}
This paper demonstrates the anisotropic response of quantum critical fluctuations with respect to the direction of the magnetic field $B$ in Ni-doped CeCoIn$_5$ by measuring the magnetization $M$ and specific heat $C$. The results show that $M/B$ at $B=0.1\ {\rm T}$ for both the tetragonal $c$ and $a$ directions exhibits $T^{-\eta}$ dependencies, and that $C/T$ at $B=0$ follows a $-\ln T$ function, which are the characteristics of non-Fermi-liquid (NFL) behaviors. For $B\,||\,c$, both the $M/B\propto T^{-\eta}$ and $C/T \propto -\ln T$ dependencies change into nearly temperature-constant behaviors by increasing $B$, indicating a crossover from the NFL state to the Fermi-liquid state. For $B\,||\,a$, however, the NFL behavior in $C/T$ persists up to $B=7\ {\rm T}$, whereas $M/B$ exhibits temperature-independent behavior for $B\ge 1\ {\rm T}$. These contrasting characteristics in $M/B$ and $C/T$ reflect the anisotropic nature of quantum critical fluctuations; the $c$-axis spin component significantly contributes to the quantum critical fluctuations. We compare this anisotropic behavior of the spin fluctuations to superconducting properties in pure CeCoIn$_5$, especially to the anisotropy in the upper critical field and the Ising-like characteristics in the spin resonance excitation, and suggest a close relationship between them. 
\end{abstract}

\maketitle

\section{Introduction}
The role of spin fluctuations in unconventional superconductivity is a long-standing subject in the physics of strongly correlated electron systems. The unconventional superconducting (SC) phase commonly emerges in the vicinity of magnetic orders in many strongly correlated electron systems, such as high-$T_c$ cuprates, FeAs-based alloys, and heavy fermion compounds. In particular, the heavy fermion compounds often exhibit SC order proximity to a magnetic quantum critical point (QCP), corresponding to a magnetic phase transition at zero temperature. Hence, quantum critical fluctuations that are enhanced around the QCP are expected to play a critical role in the SC order of the heavy fermion compounds.

Among the heavy fermion superconductors, CeCoIn$_5$ has attracted continuous interest because of its anomalous SC properties coupled with magnetic correlations \cite{rf:Petrovic2001}. This compound has a HoCoGa$_5$-type tetragonal structure [Fig.\ 1(b), inset] and exhibits a SC order below $T_c=2.3\ {\rm K}$. The magnetically mediated pairing mechanism of the SC order is inferred from the $d$-wave ($d_{x^2-y^2}$) symmetry of the SC gap \cite{rf:Izawa2001,rf:An2010,rf:Park2008}. The inelastic neutron scattering experiments have revealed that a resonance excitation involving the tetragonal $c$-axis spin component develops in the SC state \cite{rf:Stock2008,rf:Raymond2015,rf:Song2016,rf:Mazzone2017,rf:Stock2018,rf:Eremin2008,rf:Chubukov2008,rf:Michal2011}. Furthermore, applying the magnetic field $B$ yields another SC phase that coexists with an incommensurate antiferromagnetic (AFM) modulation (the so-called $Q$ phase) at very low temperatures below 0.3 K and at high fields just below $H_{c2}$ for  $B\,\perp \,c$ \cite{rf:Bianchi2003-1,rf:Kakuyanagi2005,rf:Young2007,rf:Kenzelmann2008,rf:Aperis2008,rf:Agterberg2009,rf:Yanase2008,rf:Yanase2009}. All of these features indicate a close coupling between the anisotropic spin correlations and the SC state, but the nature of the spin correlations with respect to CeCoIn$_5$ has not yet been fully uncovered.

A key to clarifying the relationship between the spin correlations and the anomalous SC properties is expected to be found in the field-induced non-Fermi-liquid (NFL) behaviors observed under $B$ when applied along the $c$ axis. At $B\sim 5\ {\rm T}$, the specific heat divided temperature exhibits $-\ln T$ dependence, and both electrical resistivity and magnetization follow nearly $T$-linear functions down to very low temperatures \cite{rf:Bianchi2003-2,rf:Paglione2003,rf:Tayama2002}. It is widely believed that spin fluctuations enhanced near an AFM QCP are responsible for these NFL behaviors \cite{rf:Paglione2003,rf:Bianchi2003-2,rf:Tokiwa2013}. In fact, substituting the ions for elements in CeCoIn$_5$, such as Nd for Ce \cite{rf:Hu2008,rf:Raymond2014}, Rh for Co \cite{rf:Zapf2001,rf:Yoko2006,rf:Yoko2008,rf:Ohira-Kawamura2007}, and Cd, Hg, and Zn for In \cite{rf:Pham2006,rf:Nicklas2007,rf:Yoko2014,rf:Yoko2015}, can induce long-range AFM orders. Moreover, possible field-induced AFM ordering at extremely low temperatures ($T\le 20\ {\rm mK}$) has been proposed by a recent quantum oscillation measurement for pure CeCoIn$_5$ \cite{rf:Shishido2018}. 

In contrast, the substitutions of Sn for In \cite{rf:Bauer2005,rf:Bauer2006,rf:Ramos2010} and Ni for Co \cite{rf:Otaka2016} do not induce the AFM phase, but simply yield paramagnetic ground states through the suppression of the SC phase. In a recent study, we have revealed that in the mixed compound CeCo$_{1-x}$Ni$_x$In$_5$, the SC transition temperature $T_c$ monotonically decreases from 2.3 ($x=0$) to 0.8 K ($x=0.20$) with increasing $x$; subsequently, the SC order disappears above the critical Ni concentration $x=0.25$ \cite{rf:Otaka2016}. At this concentration, the NFL behaviors are realized around the zero field, characterized by the $-\ln T$ dependence in the specific heat divided by the temperature, the weak diverging behavior in the magnetization, and the nearly $T$-linear behavior of the electrical resistivity \cite{rf:Otaka2016}. These NFL features are quite similar to those seen in pure CeCoIn$_5$, strongly suggesting that the NFL anomaly observed in Ni-doped CeCoIn$_5$ also originates from the AFM quantum critical fluctuations. Furthermore, the effective magnetic moment for $x\le 0.3$, estimated from the Curie-Weiss law at high temperatures, is nearly independent of $x$ and coincides well with that calculated from the $J=5/2$ multiplet in the Ce$^{3+}$ ion \cite{rf:Otaka2016}, suggesting that the Ce 4$f$ electrons are mainly responsible for the magnetic properties in pure and Ni-doped CeCoIn$_5$.  

The observation of the NFL behavior at the zero field in Ni-doped CeCoIn$_5$ provides an opportunity to investigate the magnetic anisotropy of the quantum critical fluctuations. In pure CeCoIn$_5$, in contrast, it is difficult to perform such an investigation with typical macroscopic measurements, because the quantum critical behavior is suppressed (or masked) by the SC phase at low magnetic fields and is visible only at very low temperatures above $\mu_0H_{c2}$ ($4.9\ {\rm T}$ for $B\,||\,c$ and $11.6\ {\rm T}$ for $B\,||\,a$) \cite{rf:Ronning2005,rf:Hu2012}. Consequently, the magnetic anisotropy of the quantum critical fluctuations remains unclear. In this paper we demonstrate the anisotropic changes of the NFL behaviors in the magnetization and specific heat between $B\,||\,c$ and $B\,||\,a$ in CeCo$_{1-x}$Ni$_x$In$_5$, and we discuss the relationship between the anisotropic spin fluctuations and the SC properties in pure and Ni-doped CeCoIn$_5$. 

\section{Experiment Details}
A single crystal of CeCo$_{1-x}$Ni$_{x}$In$_5$ with $x=0.25$ was grown using an Indium flux technique, the details of which are described elsewhere \cite{rf:Otaka2016}. The energy dispersive x-ray spectroscopy (EDS) and the inductively coupled plasma mass spectrometry (ICP-MS) measurements for the sample indicated that the actual Ni concentration approximately coincided with the starting (nominal) value within the deviation of $\Delta x/x\sim 17\%$, including the experimental error. Furthermore, through the EDS measurements, we confirmed the homogeneous distributions of the elements in the single crystal prepared for the experiments. The magnetization along the $c$ and $a$ axis was measured in temperatures as low as 0.1 K and in the magnetic field $B$ ($\mu_0H$) at up to 8 T with a capacitively detected Faraday force magnetometer \cite{rf:Sakakibara94}. A commercial SQUID magnetometer (MPMS, Quantum Design) was used for the magnetization measurements in the temperature range of 2.0--300 K and the magnetic field at up to 5 T. The specific heat $C_p$ was measured in the temperature range of 0.31--4 K and in the field range of 0--7 T with a conventional quasiadiabatic technique.

\section{Results}
Figures 1(a) and 1(b) show the temperature dependencies of the $c$- and $a$-axis magnetization divided by the magnetic field $M/B$ respectively. Note that the $M/B$ data are plotted with logarithmic scales for both the vertical and horizontal axes. $M/B$ for both directions showed qualitatively similar features. Namely, $M/B$ at the lowest field ($B=0.1\ {\rm T}$) exhibited diverging behavior with a $T^{-\eta}$ function ($\eta < 1$) as the temperature decreased. The $T^{-\eta}$ dependence in $M/B$ was realized in a very wide temperature range of 0.1--10 K for both directions. It is natural to conclude that this NFL behavior originates from the quantum critical fluctuations, because similar NFL behaviors are also found in various macroscopic quantities in pure CeCoIn$_5$ \cite{rf:Bianchi2003-2,rf:Paglione2003,rf:Tokiwa2013} and its doped alloys \cite{rf:Bauer2005,rf:Bauer2006,rf:Yoko2017}. In both the $c$- and $a$-axis magnetization, the diverging feature was reduced by further applying $B$, and the $M/B$-constant behavior was then realized at low temperatures.

\begin{figure}[tbp]
\begin{center}
\includegraphics[bb=0 0 472 539,keepaspectratio,width=0.45\textwidth]{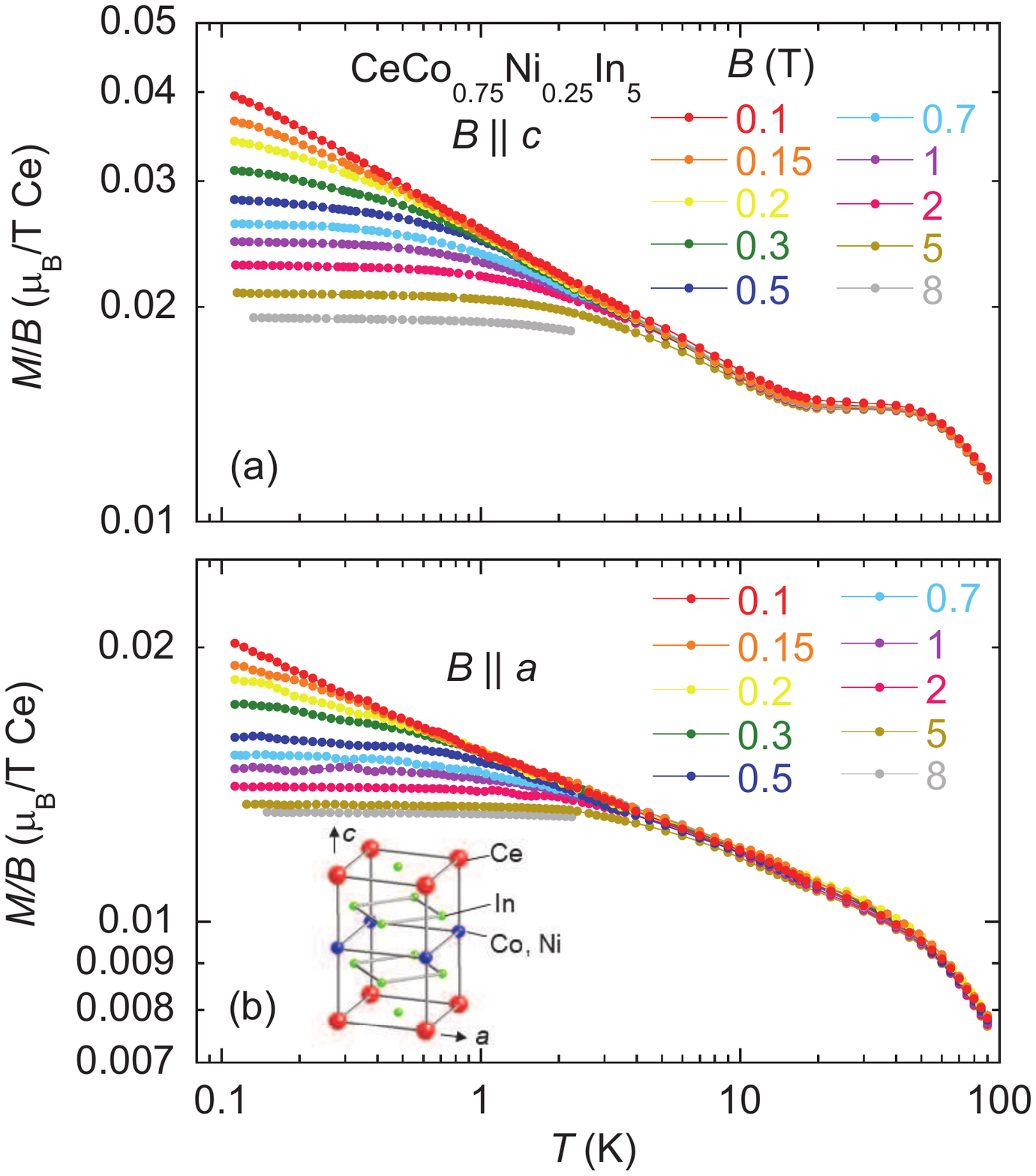}
\end{center}
  \caption{
Temperature variations of (a) the $c$-axis magnetization and (b) the $a$-axis magnetization divided by the magnetic field for CeCo$_{0.75}$Ni$_{0.25}$In$_5$. In (a) and (b), logarithmic scales are used for both the vertical and horizontal axes. The crystal structure of CeCo$_{1-x}$Ni$_{x}$In$_5$ is depicted in the inset of (b). 
}
\end{figure}

From a quantitative viewpoint, however, a significant anisotropy was found in the NFL region between the $c$-axis and $a$-axis magnetization $M_c$ and $M_a$, respectively. The exponent $\eta$ of $M/B\propto T^{-\eta}$ at $B=0.1\ {\rm T}$ in $M_c$ [$\eta_c=0.20(2)$] was larger than the value [$\eta_a=0.12(2)$] in $M_a$, as the details of those evaluation procedures are described later. Furthermore, the magnitude of $M_c$ at $B=0.1\ {\rm T}$ and $T=0.1\ {\rm K}$ was twice that of $M_a$. The diverging behavior in the temperature variation of $M_c$ was thus stronger than that of $M_a$. Indeed, this feature can be verified by considering the variation of $M_c/M_a$ as a function of temperature (Fig.\ 2). $M_c/M_a$ exhibited a peak with a magnitude of $1.5$ at $\sim 60\ {\rm K}$. The peak structure was also observed in the other physical quantities, such as the electrical resistivity \cite{rf:Petrovic2001}, and its origin is considered a development of a coherent heavy-fermion state below this temperature. The $M_c/M_a$ value was reduced to $1.32$ with decreasing temperatures, down to $\sim 15\ {\rm K}$. However, the spin fluctuations, associated with the NFL behavior, enhanced the $M_c/M_a$ value at low temperatures again; $M_c/M_a$ for $B=0.1\ {\rm T}$ increased with a decrease in temperature below $\sim 15\ {\rm K}$ and then reached 1.96(10) at 0.11 K. In a high magnetic-field region, in contrast, $M_c/M_a$ for $B=5$ and 8 T exhibited a saturation to the values of 1.55(3) and 1.46(2) at low temperatures, respectively, remaining with magnitudes comparable to those in high temperatures. These experimental results suggest that the NFL anomaly involved mainly the $c$-axis spin component.
   
\begin{figure}[tbp]
\begin{center}
\includegraphics[bb=0 0 471 338,keepaspectratio,width=0.45\textwidth]{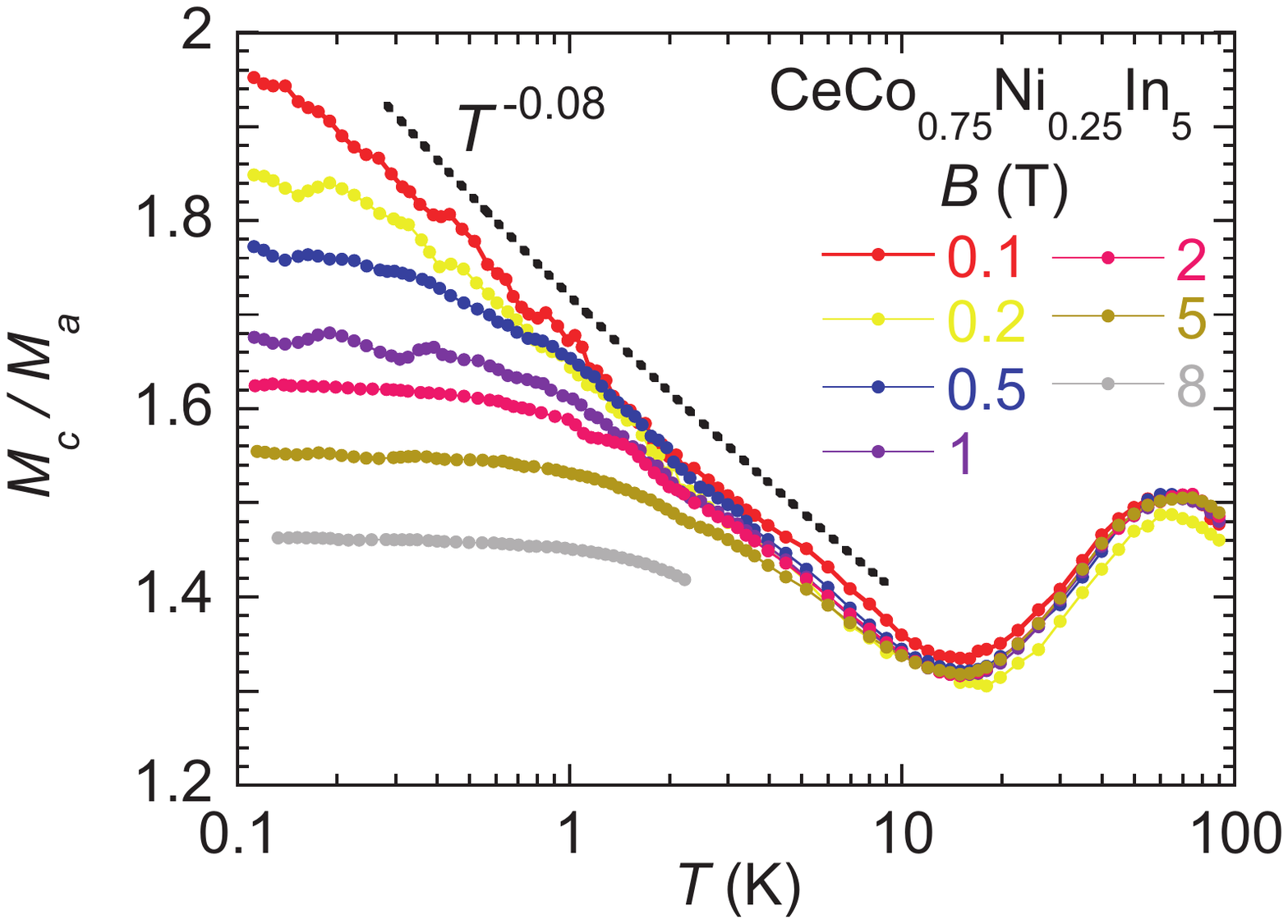}
\end{center}
  \caption{
Temperature variations of the ratio of the $c$-axis and the $a$-axis magnetization $M_c/M_a$ for CeCo$_{0.75}$Ni$_{0.25}$In$_5$. A $T^{-0.08}$ $[=T^{-(\eta_c-\eta_a)}]$ function is represented as a dashed line for comparison.
}
\end{figure}

Figures 3(a) and 3(b) show the specific heat divided by the temperature $C_p/T$ obtained under various fields along the $c$ and $a$ axis, respectively. For $B\,||\,c$, $C_p/T$ was markedly enhanced below $\sim 0.7\ {\rm K}$ for $B \ge 3\ {\rm T}$, although its temperature dependence became weak with increasing $B$ at high temperatures. This enhancement is considered to be caused by the Zeeman splitting of the nuclear spins. Such an effect should also be included in $C_p/T$ for $B\,||\,a$. To eliminate this contribution, we estimated the nuclear Schottky anomaly $C_{\rm nucl}$ by performing a calculation based on the natural abundance of the nuclear spins in the sample [Fig.\ 3(b), inset]. At $B=7\ {\rm T}$ and 0.4 K, the fraction of $C_{\rm nucl}$ in $C_p$ was estimated to be 14\% for $B\,||\,c$ and 12\% for $B\,||\,a$. Note that the contribution of Ni and Co nuclear spins was only 10\% in $C_{\rm nucl}$; therefore, the ambiguity of the Ni/Co concentration ($\Delta x/x\sim 17\%$) in the sample is negligible in the estimation of $C_{\rm nucl}$.  
\begin{figure}[tbp]
\begin{center}
\includegraphics[bb=0 0 456 461,keepaspectratio,width=0.45\textwidth]{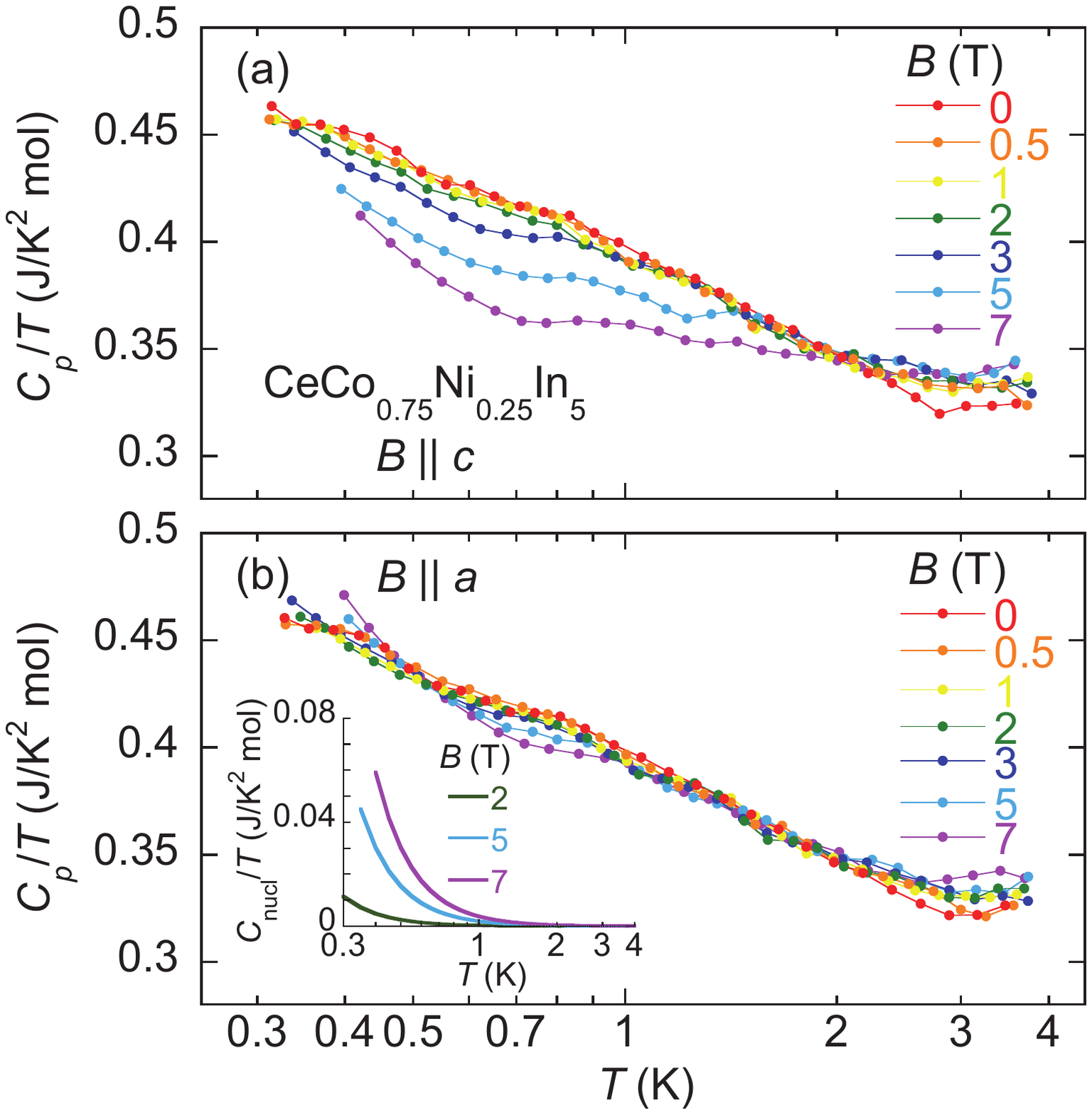}
\end{center}
  \caption{
Temperature variations of specific heat divided by the temperature $C_p/T$ for CeCo$_{0.75}$Ni$_{0.25}$In$_5$, measured under various $B$ with the directions of (a) $B\,||\,c$ and (b) $B\,||\,a$. The inset of (b) shows the nuclear Schottky contribution $C_{\rm nucl}/T$ calculated based on the natural abundance of nuclear spins in the sample.
}
\end{figure}

Figures 4(a) and 4(b) display the specific heat data obtained by subtracting the nuclear spin contribution $C/T \equiv (C_p -C_{\rm nucl})/T$ for $B\,||\,c$ and $B\,||\,a$, respectively. $C/T$ for $B=0$ increased with decreasing temperature, with a nearly $-\ln T$ dependence at temperatures as low as 0.31 K. As displayed in Fig.\ 4(a), this feature was markedly suppressed by applying $B$ along the $c$ axis, and $C/T$ eventually became nearly independent of temperature at $B=7\ {\rm T}$. The feature of suppression in $C/T$ coincides fairly well with that observed in $M_c/B$ [Fig.\ 1(a)]; hence, these behaviors are attributed to a crossover from the NFL to Fermi-liquid (FL) states.
\begin{figure}[tbp]
\begin{center}
\includegraphics[bb=0 0 457 458,keepaspectratio,width=0.45\textwidth]{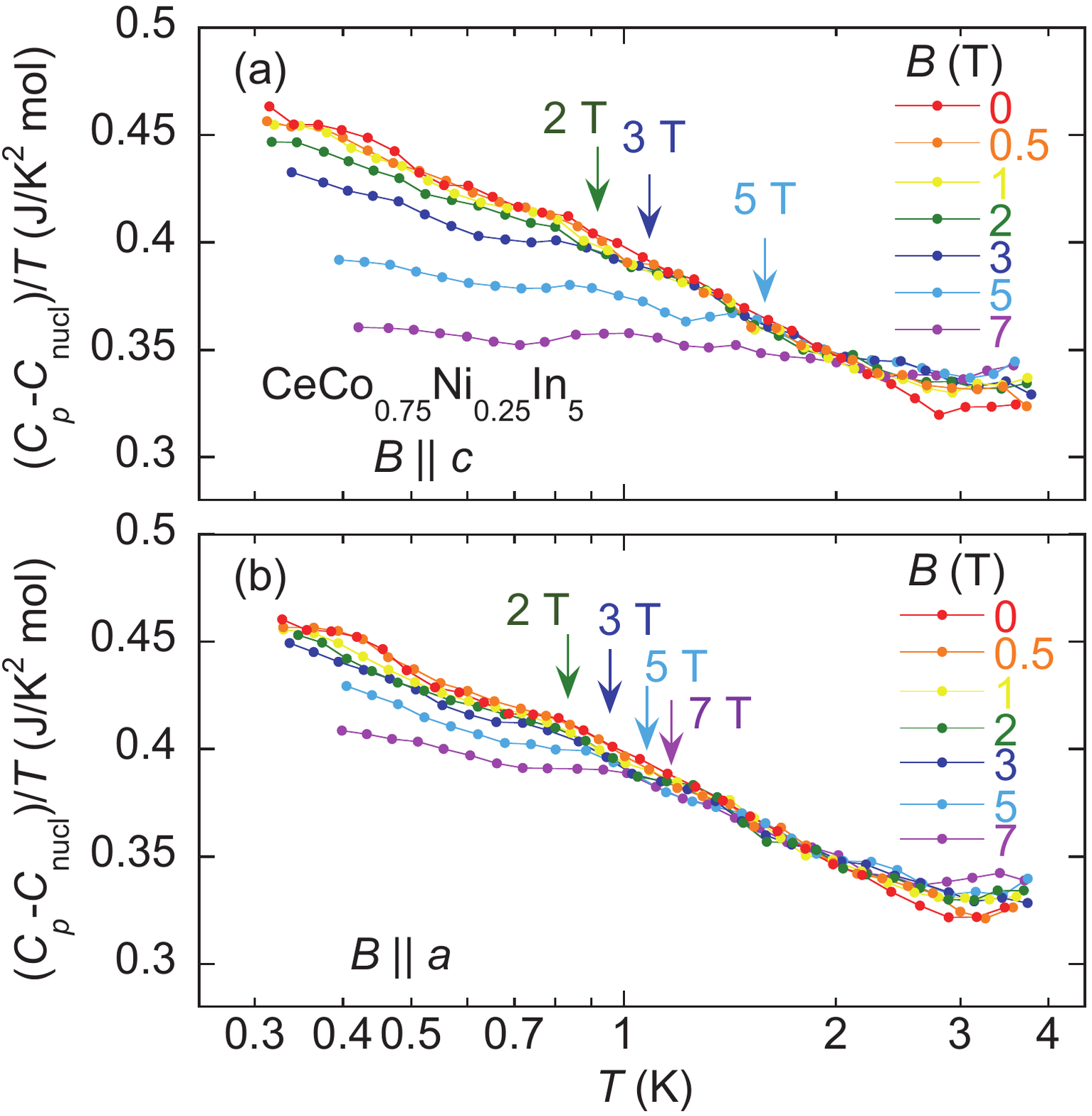}
\end{center}
  \caption{
Low temperature specific heat divided by the temperature obtained by subtracting the nuclear spin contribution $C/T \equiv (C_p -C_{\rm nucl})/T$ for CeCo$_{0.75}$Ni$_{0.25}$In$_5$ under $B$ with (a) $B\,||\,c$ and (b) $B\,||\,a$. The arrows indicate the characteristic temperature $T^*$ below which $C/T$ deviates from the $-\ln T$ function.
}
\end{figure}

However, it was found that the $-\ln T$ diverging behavior in $C/T$ was not suppressed as much by $B$ for $B\,||\,a$. At 0.4 K, the reduction of the specific heat at $B=5\ {\rm T}$, $1-C(5\ {\rm T})/C(0\ {\rm T})$, was estimated to be 5\% for $B\,||\,a$, whereas it was 14\% for $B\,||\,c$. In addition, $C/T$ at $B=7\ {\rm T}$ continued to increase with the decreasing temperature for $B\,||\,a$, whereas it was nearly independent of temperature for $B\,||\,c$. Similar weak $B$ dependence of $C/T$ for $B\,||\,a$ was also found at a very high $B$ region above $\mu_0H_{c2}=11.6\ {\rm T}$ in pure CeCoIn$_5$ \cite{rf:Ronning2005}. This weak $B$ dependence in $C/T$ for $B\,||\,a$ is in stark contrast to the rapid reduction of $M_a/B$ with $B$ for the same $B$ direction; $M_a/B$ was markedly suppressed by applying $B$ and then became constant at low temperatures for $B\ge 1\ {\rm T}$ [Fig.\ 1(b)]. These contrasting features in $C/T$ and $M_a/B$ for $B\,||\,a$ strongly suggest that the fluctuating spin component is perpendicular to the applied $B$ direction; that is, the $c$-axis spin component significantly contributes to the quantum critical fluctuations in CeCo$_{1-x}$Ni$_x$In$_5$. This situation is similar to that expected in the Ising model with a transverse magnetic field, in which the transverse magnetic field does not align the spins but yields a quantum paramagnetic state with short-range spin correlations \cite{rf:Sachdev99}. However, it should be remembered that the anisotropy of magnetic moments in the present system was not so strong that it can be regarded as simply the Ising-like anisotropy. In addition, the spins of the itinerant heavy quasiparticles, rather than the completely localized spins, were likely responsible for the quantum critical fluctuations. Hence, the deviation from the Ising-like characteristics of the magnetic moments would lead to a suppression of the quantum critical fluctuations and would then stabilize the FL state at a high $B$ region above $B=7\ {\rm T}$, even for $B\, ||\,a$.

In Figs.\ 5(a) and 5(b) we summarize the exponent $\eta$ of $M/B\propto T^{-\eta}$ at low temperatures for $B\,||\,c$ and $B\,||\,a$, respectively. In these plots, $\eta$ was estimated using a simple relation: $\eta=-T/(M/B)\,d(M/B)/dT$. The effect of the Van Vleck susceptibility $\chi_V$ may be included in the estimation of $\eta$ using an alternative formula: $\eta=-T/(M/B-\chi_V)\,d(M/B)/dT$. However, we confirmed that the trends seen in Figs.\ 5(a) and 5(b) did not depend on the finite $\chi_V$ value, at least up to $\chi_V \sim 2.5\times 10^{-3} \ \mu_B$/T Ce, which is about half of the magnitude of the experimentally observed magnetic susceptibility at 300 K \cite{rf:Otaka2016}.

\begin{figure}[tbp]
\begin{center}
\includegraphics[bb=0 0 562 657,keepaspectratio,width=0.43\textwidth]{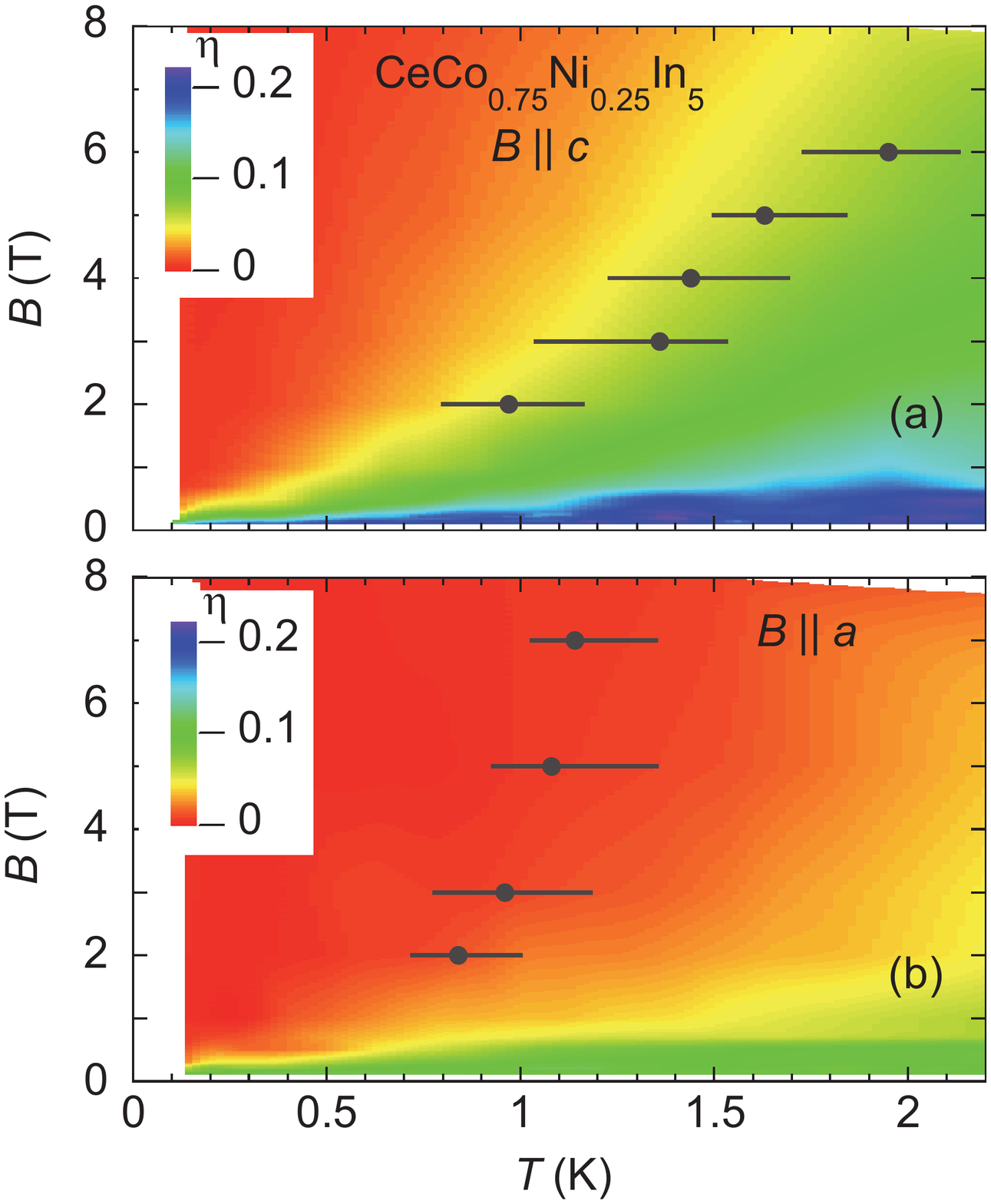}
\end{center}
  \caption{
The image plots of the exponent $\eta$ in $M/B\propto T^{-\eta}$ for CeCo$_{0.75}$Ni$_{0.25}$In$_5$, depicted in the magnetic field versus the temperature plane for (a) $B\,||\,c$ and (b) $B\,||\,a$. The filled circles indicate the characteristic temperature $T^*$ of $C/T$, below which $C/T$ deviates from the $-\ln T$ function.
}
\end{figure}

As displayed in Fig.\ 5(a), the NFL state with $\eta = 0.2$ governed the low $B$ region in the $B-T$ plane for $B\,||\,c$, and the suppression of the NFL state at the high $B$ region was realized as a reduction of  $\eta$ toward $\sim 0$. A similar gradual suppression of the NFL behavior with $B$ was also observed in $C/T$. In fact, when the characteristic temperature $T^*$ below which $C/T$ deviates from the $-\ln T$ function [see Fig.\ 4(a)] was plotted onto the image map of $\eta$ in Fig.\ 5(a), we found that the $T^*(B)$ curve traced the contour of $\eta=0.06$ well. This consistency between $M/B$ and $C/T$ reflected the occurrence of the NFL-to-FL crossover for $B\,||\,c$. 

However, the situation for $B\,||\,a$ was very different, as shown in Fig.\ 5(b). The finite $\eta$ value for $B\sim 0$ was rapidly suppressed by $B$, and a large $\eta\sim 0$ region was then distributed in the $B-T$ plane. In contrast, the $T^*(B)$ curve entered deeply into the $\eta\sim 0$ region. Note that for $B\,||\,a$, $C/T$ did not exhibit the $C/T$-constant behavior ascribed to the FL state, even below $T^*$, although $T^*$ could be defined in the $C/T$ data. It is likely that the $c$-axis spin component of the quantum critical fluctuations, which was not significantly influenced by $B$ for $B\,||\,a$, led to this inconsistency between $M/B$ and $C/T$ for $B\,||\,a$, as argued previously. 

\section{Discussion}
The present investigation of CeCo$_{0.75}$Ni$_{0.25}$In$_5$ revealed the clear anisotropic response of the NFL behaviors in $M/B$ and $C/T$ with respect to the $B$ direction; the crossover from the NFL state to the FL state occurred for $B\,||\,c$, whereas the NFL behavior persisted at least up to $B=7\ {\rm T}$ for $B\,||\,a$. In this section, we compare this anisotropic NFL behavior to the SC properties in pure CeCoIn$_5$. 

First, we find that the anisotropy concerning the stability of the SC phase in pure CeCoIn$_5$ qualitatively coincides with that of the quantum critical fluctuations in Ni25\%-doped CeCoIn$_5$. The SC order parameter in the pure compound has a characteristic temperature scale of $\sim 2\ {\rm K}$, corresponding to $T_c$. In addition, this SC state is broken at $\mu_0H_{c2}=4.9\ {\rm T}$ for $B\,||\,c$, although it persists up to $11.6\ {\rm T}$ for $B\,||\,a$ \cite{rf:Ikeda2001}. In the Ni25\%-doped alloy, $C/T$ at $\sim 2\ {\rm K}$ takes on the NFL characteristics up to the same $B$ ranges as $H_{c2}$ in the pure compound. This coincidence of the stability of the NFL and SC states concerning $B$ implies that the spin correlations, similar to those yielding the NFL behavior in the Ni25\%-doped alloy, play a critical role in the occurrence of the SC order in the pure compound. If this is the case, such spin correlations should be concerned with the determination of $H_{c2}$ through both the SC condensation energy and the paramagnetic spin susceptibility, yielding a Pauli paramagnetic effect \cite{rf:Yanase2008}, because $H_{c2}$ of CeCoIn$_5$ is considered Pauli limited \cite{rf:Izawa2001,rf:Ikeda2001}.

Second, it is remarkable that the $c$-axis spin component is primarily responsible for the quantum critical fluctuations in Ni25\%-doped CeCoIn$_5$. Indeed, such anisotropic spin fluctuations and excitations are also observed in the SC phase of pure CeCoIn$_5$. The recent inelastic neutron scattering experiments for pure and Nd-doped CeCoIn$_5$ have revealed that the spin resonance excitation emerging in the SC phase has a nearly Ising nature along the $c$ axis \cite{rf:Raymond2015,rf:Mazzone2017}. This similarity in the spin polarization suggests that the spin resonance excitation in pure CeCoIn$_5$ and the NFL behavior in Ni-doped CeCoIn$_5$ have similar origins. The spin fluctuations in Ni25\%-doped CeCoIn$_5$ may have an energy distribution centered at $\hbar\omega\sim 0$, because the NFL behavior at $B\sim 0$ in $M/B$ and $C/T$ persists down to very low temperatures. However, once the SC order occurs, as in pure CeCoIn$_5$, the spin fluctuations may have gapped energy due to the SC condensation, detected as the spin resonance excitation in the inelastic neutron scattering measurements. In this situation, the coherency and Ising-like polarization of the spin fluctuations may be somewhat enhanced along with the variation of the ground state from the paramagnetic NFL state to the SC ordered phase. In fact, it has been demonstrated that the spin resonance excitation may condensate into AFM ordering \cite{rf:Song2016,rf:Stock2018,rf:Michal2011}, supporting the above suggestion that the quantum critical fluctuations and the spin resonance excitation have similar origins because the quantum critical fluctuations likely originate from the AFM instabilities in pure CeCoIn$_5$ and its doped alloys \cite{rf:Paglione2003,rf:Pham2006,rf:Yoko2017}. 

Despite the aforementioned considerations, the microscopic nature of the quantum critical fluctuations has not yet been uncovered. We believe that the relationship of the spin fluctuations between pure and Ni25\%-doped CeCoIn$_5$ would be clarified by comprehensive investigations using the inelastic neutron scattering technique on CeCo$_{1-x}$Ni$_{x}$In$_5$ with a wide $x$ range. Such investigations could provide a key to understanding the anomalous SC properties coupled with the magnetic correlations in CeCoIn$_5$. 

\section{Conclusion}
Our magnetization and specific heat measurements for CeCo$_{0.75}$Ni$_{0.25}$In$_5$ revealed anisotropic NFL behavior, depending on the $B$ direction. For $B\,||\,c$, the diverging behaviors in the temperature variations of $M/B$ and $C/T$ changed into nearly $T$-constant behaviors, reflecting the NFL-to-FL crossover with increasing $B$. For $B\,||\,a$, however, the NFL behavior in $C/T$ persisted up to $B=7\ {\rm T}$, although $M/B$ was sufficiently reduced with $B$ for $B \ge 1\ {\rm T}$. These anisotropic responses in $M/B$ and $C/T$ indicate that the quantum critical fluctuations are suppressed by the $c$-axis magnetic field more effectively than by the $a$-axis field because they are composed mainly of the $c$-axis spin component. We compared this feature to the SC properties in pure CeCoIn$_5$, especially to the anisotropy in the upper critical field and the Ising-like characteristics in the spin resonance excitation, and suggested a close coupling between them.

\begin{acknowledgments}
We are grateful to Y.\ Oshima, I.\ Kawasaki, R.\ Otaka, and Rahmanto for their experimental support, and to D.\  Ueta and T.\ Masuda for their assistance with the specific-heat measurements prior to this study. M.Y. expresses gratitude to Y.\ Yanase for fruitful discussions. This study was supported in part by JSPS KAKENHI Grant Number 17K05529, and the research completed at ISSP was supported by a Grant-in-Aid for Scientific Research on Innovative Areas ``J-Physics" (15H05883) from JSPS.
\end{acknowledgments}

\end{document}